\documentclass[aps,prd,preprint,superscriptaddress,showpacs,floatfix,nobibnotes]{revtex4-1}

\usepackage[latin1]{inputenc}
\usepackage{comment}
\usepackage{graphicx}
\usepackage{setspace}
\usepackage{amsmath}
\usepackage{geometry}   
\usepackage{amsfonts}
\usepackage{amssymb}
\usepackage{amsthm}
\usepackage{slashed}
\usepackage{esint}
\usepackage{mathrsfs}
\usepackage{enumerate}
\usepackage{color}
\usepackage{array}

\usepackage[yyyymmdd,hhmmss]{datetime}

\numberwithin{equation}{section}

\newcommand{\bea}{\begin{eqnarray}}
\newcommand{\eea}{\end{eqnarray}}

\newcommand{\BbbR}{\mathbb{R}}

\newcommand{\BbbN}{\mathbb{N}}

\theoremstyle{plain}
\newtheorem{thm}{Theorem}[section]
\newtheorem{lem}[thm]{Lemma}

\newtheorem{cond}[thm]{Conditions}

\begin{document}
\title{Smooth and sharp creation of a spherical shell \\ 
for a $(3+1)$-dimensional quantum field}

\author{Margaret E. Carrington}
\email[]{carrington@brandonu.ca} 
\affiliation{Department of Physics, Brandon University, Brandon, Manitoba, R7A 6A9 Canada}
\affiliation{Winnipeg Institute for Theoretical Physics, Winnipeg, Manitoba}

\author{Gabor Kunstatter}
\email[]{gkunstatter@uwinnipeg.ca} 
\affiliation{Winnipeg Institute for Theoretical Physics, Winnipeg, Manitoba}
\affiliation{Department of Physics, University of Winnipeg, Winnipeg, Manitoba, R3B 2E9 Canada}

\author{Jorma Louko}
\email[]{jorma.louko@nottingham.ac.uk} 
\affiliation{School of Mathematical Sciences, University of Nottingham, Nottingham NG7 2RD, United Kingdom}

\author{L.~J. Zhou}
\email[]{zhoulingjunjeff@gmail.com} 
\affiliation{Department of Physics, Brandon University, Brandon, Manitoba, R7A 6A9 Canada}
\affiliation{Winnipeg Institute for Theoretical Physics, Winnipeg, Manitoba}
\affiliation{Department of Physics, University of Winnipeg, Winnipeg, Manitoba, R3B 2E9 Canada}
\affiliation{Department of Physics and Astronomy, University of Manitoba, Winnipeg, Manitoba, R3T 2N2 Canada}

\date{\today}

\begin{abstract}
We study the creation of a spherical, finite radius source for a quantized massless scalar field in 3+1 dimensions. The goal is to model the breakdown of correlations that has been proposed to occur at the horizon of an evaporating black hole. We do this by introducing at fixed radius $r=a$ a one parameter family of self-adjoint extensions of the three dimensional Laplacian operator that interpolate between
the condition that the
values and the derivatives on the two
sides of $r=a$ coincide 
for $t\le0$ (no wall) and the two-sided Dirichlet boundary condition for $t \ge 1/\lambda$ (fully-developed wall).
Creation of the shell produces null, spherical pulses of energy on either side of the shell, one ingoing and the other outgoing. 
The renormalized energy density $\langle T_{00}\rangle$ diverges 
to positive infinity in the outgoing energy pulse, just outside the light cone of the fully-formed wall at $t=1/\lambda$. 
Unlike in the 3+1 point source creation, there is no persistent memory cloud of energy. As in the creation of a 1+1 dimensional wall, the response of an Unruh-DeWitt detector 
in the post-shell region 
is independent of the time scale for shell formation and is finite. 
The latter property casts doubt on the efficacy of this mechanism for firewall creation.
\end{abstract}


\normalsize
\maketitle
\tableofcontents
\section{Introduction}
It has been more than thirty years since Hawking first suggested~\cite{Hawking-on-info-loss} that black hole formation might give rise to a fundamental breakdown of predictability.  Many approaches have been taken in order to resolve this apparent dilemma \cite{reviews}. Among these is the suggestion \cite{Mathur-on-info-paradox1,Mathur-on-info-paradox2,braunstein-et-al,Almheiri:2013hfa} that the horizon of a radiating black hole might be more singular than suggested by standard quantum field theory on a curved background
\cite{birrell-davies, wald-smallbook}.  Specifically, there might exist an ``energetic curtain''~\cite{braunstein-et-al} or ``firewall''~\cite{Almheiri:2013hfa} whose purpose is to break correlations between objects falling into the horizon and those remaining on the outside. An important question then is whether there exists a local mechanism for creating such a firewall, one that does not rely on a detailed knowledge of the underlying theory of quantum gravity. 

Recently Brown and Louko \cite{brown} explored  a 1+1 dimensional model of firewall creation based on the imposition of time dependent boundary conditions at a fixed point in space.  These boundary conditions were equivalent to the insertion of a wall that broke correlations between the quantum field on the left side of the wall and that on the right. They found that in the rapid creation limit this scenario resulted in the creation of a divergent null pulse of energy (the firewall) emanating from the point at which the boundary conditions were imposed. However, the response of an Unruh-DeWitt detector crossing this pulsed remained finite, irrespective of how rapidly the wall was created, suggesting that this mechanism could not produce sufficient energy to break all correlations as required. 

A potentially more realistic 3+1 dimensional model was studied in \cite{Zhou}, where time dependent boundary conditions interpolating between Neumann-type (ordinary Minkowski dynamics) and Dirichlet-type were imposed at the spatial origin. This corresponds to the smooth creation of a point source, as opposed to a wall separating two regions of space. As one might expect from standard quantum field theory, the field's energy and the detector response are more divergent in 3+1 dimensions than in 1+1 dimensions. The 3+1 renormalized energy density  $\langle T_{00}\rangle$ was shown to be well defined everywhere away from the source but unbounded both above (after) and below (before) the energetic pulse. Moreover, in this model a cloud of positive energy lingers near the source after the source is fully formed. The total energy of this cloud is positive infinity.  At fixed radius, $r$, $\langle T_{00}\rangle$ is not static and diverges as $t\to \infty$. In the limit of rapid source creation  $\langle T_{00}\rangle$ diverges everywhere in the timelike future of the creation event. The response of an Unruh-DeWitt detector traversing the shell is divergent as desired for the firewall mechanism, but the divergence appears to be primarily due to the energetic cloud that surrounds the source.

The purpose of the present work is to extend the study of this mechanism further by considering the creation of a spherical wall in 3+1 dimensions. We do this by introducing at fixed radius $r=a$ a one parameter family of self-adjoint extensions of the three dimensional Laplacian operator that interpolate between the condition that the
values and the derivatives on the two
sides of $r=a$ coincide 
for $t\le0$ (no wall) and the two-sided Dirichlet boundary condition for $t \ge 1/\lambda$ (fully-developed wall).
As in the previous studies, wall creation produces a null pulse of energy on either side of the shell, one ingoing and the other outgoing. We find that  $\langle T_{00}\rangle$ diverges 
to positive infinity in the outgoing energy pulse, just outside the light cone of the fully-formed wall at $t=1/\lambda$. 
Unlike in
the 3+1 point source creation, there is no persistent memory cloud of energy. 
As in the 1+1 dimensional wall, the response of the detector 
in the post-shell region
is independent of the time scale $1/\lambda$ and finite, 
once again casting doubt on the efficacy of this mechanism for breaking all entanglement at black hole horizons.

The paper is organized as follows. In section \ref{mode-fcns-section}, we set up the massless Klein-Gordon equation with the boundary conditions at $r=a$ required for shell formation.  We solve for the mode functions for time $t<a$ and quantize the field. We restrict to $t<a$ in order to ensure that the ingoing pulse does not have time to reach the origin and re-disperse, thereby simplifying the calculation significantly. 
In section \ref{section-energy}, we discuss the quantized total energy with focus on the energy density in the  regions to the `future' of the outgoing pulses. A discussion of the energy density in the intermediate regions can be found in appendix \ref{outside-region}.
Section \ref{sec:detector} investigates the response of an Unruh-DeWitt detector. Section \ref{summary} concludes with a summary of the results and conclusions. Technical details are given in  five appendices.

\section{Quantization}
\label{mode-fcns-section}

We consider a real massless scalar field $\phi$ in $(3+1)$-dimensional Minkowski spacetime with field equation  
\begin{align}
(\partial^2_t-\nabla^2)\phi = 0\,.
\label{eq:fieldeq}
\end{align}
In spherical coordinates, the Laplacian is
\begin{equation}
\nabla^2=\frac{1}{r^{2}}\partial_r(r^{2}\partial_r)+\frac{1}{r^2}\nabla^2_{S^2}
\end{equation}
where $\nabla^2_{S^2}$ is the Laplacian on the unit $S^{2}$ sphere. 
We consider only the spherically symmetric sector and therefore we define
\begin{align}
\phi = \frac{f(t,r)}{\sqrt{4\pi} \; r}
\,, 
\label{eq:phi-scaledto-f}
\end{align} 
so that the field equation (\ref{eq:fieldeq}) becomes 
\begin{equation}
\big(\partial_t^2 - \partial^2_r\big) f(t,r) = 0\,.
\end{equation}

We want to consider the formation of a wall, or spherical shell, at position $r=a$ between times $t=0$ and $t=1/\lambda$. 
In order to do this, we replace the Laplacian with a one parameter family of self-adjoint extensions defined on $L_2(\mathbb{R}^3)$ with the sphere at $r=a$ removed. 
Some details are given in Appendix \ref{app:laplaceoperator}; we summarize the important points below. 
The self-adjoint extensions of the Laplacian are parametrized by the function $\theta(t)$, and we assume $\theta\in[0,\pi/2]$ so that the spectrum of the self adjoint extensions of the Klein-Gordon equation has no tachyonic modes. 
The angle $\theta(t)$ can be written in terms of a function $h$ of a dimensionless variable $T=\lambda t$ which is defined by the equation  
\bea
\theta(t) = \cot^{-1}\big[L \lambda \cot\big(h(\lambda t)\big)\big]\,.
\eea
$L$ is a positive constant of dimension length which is introduced for convenience; its length is considered fixed. 
The real solutions of the Klein-Gordon operator satisfy the boundary conditions 
\bea
\label{bc0}
&& f(t,0)=0 \\
\label{bc1}
&&f(t,a^-)=f(t,a^+)\,, \\[2mm]
\label{bc2}
&& \frac{f^\prime(t,a^+)}{f(t,a^+)}-\frac{f^\prime(t,a^-)}{f(t,a^-)} = 2\lambda \cot \big(h(\lambda t)\big)\,.
\eea

In order to model the creation of a shell at $r=a$ we choose a smooth function $h(T)$ that interpolates between $h(0)=\pi/2$ and $h(1)=0$
\begin{subequations}
\label{eq:theta-scaled}
\begin{alignat}{2}
& h(T) = \pi/2 &\quad& \text{for $T\le0$}
\ ,  
\\
& 0 < h(T) < \pi/2 && \text{for $0 < T < 1$}
\ , 
\\
& h(T) = 0 && \text{for $T\ge1$}
\ . 
\end{alignat}
\end{subequations}
From equations (\ref{bc1}, \ref{bc2}, \ref{eq:theta-scaled}) we see that the boundary condition at $r=a$ evolves from the condition at $t=T=0$ that the derivatives on the two sides of $r=a$ coincide, to Dirichlet at $\lambda t=T=1$, which can be thought of as the creation of a wall at $r=a$. In order to show that this wall has a physical interpretation which can be related to the firewall scenario, we would have to show that it produces a pulse of energy, as explained in section \ref{section-energy}. 

We will quantize the scalar field by writing
\bea
\phi(t,r) = 
\int^{\infty}_{0} \bigl( a_k \phi_k(t,r)+a^\dagger_k \overline{\phi}_k(t,r) \bigr) \, dk\,,
\label{eq:phi-op}
\eea
or equivalently 
\bea
f(t,r) = \int^{\infty}_{0} \bigl( a_k U_k(t,r)+a^\dagger_k \overline{U}_k(t,r) \bigr) \, dk \,,
\label{eq:f-op}
\eea
where 
\bea
U_k(t,r) = \frac{\phi_k(t,r)}{\sqrt{4\pi}r}\,,
\label{Ukdef}
\eea
and the annihilation and creation operators have the commutators 
$\bigl[a_k,a^\dagger_{k'}\bigr] = \delta(k-k')$. 
The mode functions $U_k(t,r)$ will be normalized so that the field and its time derivative have the correct equal-time commutator. 
The vacuum $|0\rangle$ is the state that is annihilated by all 
$a_k$.
%
We work in the radial null coordinates $u := t-r$ and $v := t+r$, 
and write the field equation 
\eqref{eq:fieldeq}  
\begin{align}
\partial_u \partial_v f =0 
\,.   
\end{align}

We construct an ansatz for the mode functions in two regions inside ($r<a$) and outside ($r>a$) the shell. 
We consider the case when $a>1/\lambda$ which means that (in units where $c=1$) the timescale for wall formation is less than the distance between the location of the wall and the origin. Physically this means that during the wall formation process only left-movers are modified inside the shell, and only right-movers are modified outside the shell. Our ansatz is
\begin{align}
\label{anz}
U_k(t,r) = 
\begin{cases}
\frac{1}{\sqrt{4\pi k}} \big(e^{-i k v} + E_k(u) \big)  & \text{for $r>a$}
\ , 
\\
\frac{1}{\sqrt{4\pi k}} \big(G_k(v) -e^{-i k u}  \big)  & \text{for $0<r<a$}\,.
\end{cases}
\end{align}
This form of $U_k$ satisfies the Klein-Gordon equation for any choice of the functions $G_k(u)$ and $E_k(v)$. Our goal is to find a solution for these functions so that the boundary conditions (\ref{bc0}, \ref{bc1}, \ref{bc2}) are satisfied. Substituting (\ref{anz}) into (\ref{bc1}, \ref{bc2}) we obtain
\begin{subequations}
\label{bc3}
\begin{alignat}{2}
& -e^{-ik(t-a)}+G_k(t+a)=e^{-ik(t+a)}+E_k(t-a)\\[2mm]
& 2\lambda\cot h(\lambda t)= \frac{-ike^{-ik(t+a)}-E'_k(t-a)}{e^{-ik(t+a)}+E_k(t-a)}-\frac{-ike^{-ik(t-a)}+G'_k(t+a)}{-e^{-ik(t-a)}+G_k(t+a)}
\end{alignat}
\end{subequations}
where the prime indicates differentiation with respect to the argument.  
To solve \eqref{bc3}, we use the dimensionless time variable $T=\lambda t$ introduced previously and define the auxiliary function $B(T)$ 
\begin{align}
B(T) = 
\begin{cases}
{\displaystyle{1}}
& \text{for $T \le 0$}
\ , 
\\
{\displaystyle{\exp\left(\int_{0}^{T}\cot\bigl(h(z)\bigr) \, dz \right)}}
& \text{for $0 < T < 1$}
\ . 
\end{cases}
\label{eq:B-sol}
\end{align}
From (\ref{eq:theta-scaled}) it is easy to see that $B(0)=1$ and that for $0\le T<1$, $B(T)$ is smooth and  satisfies 
\begin{align}
\frac{B'(T)}{B(T)}
= \cot\bigl(h(T)\bigr) 
\,.
\label{eq:Bscaled-diffeq}
\end{align}
In Appendix B we show that $1/B(T)$ and all of its derivatives approach zero as $T\to 1_-$, and therefore $1/B(T)$ is smooth at $T=1$, but $B(T) \to \infty$ as $T\to 1_-$.

\begin{figure}[h]
\begin{center}
\includegraphics[width=10cm]{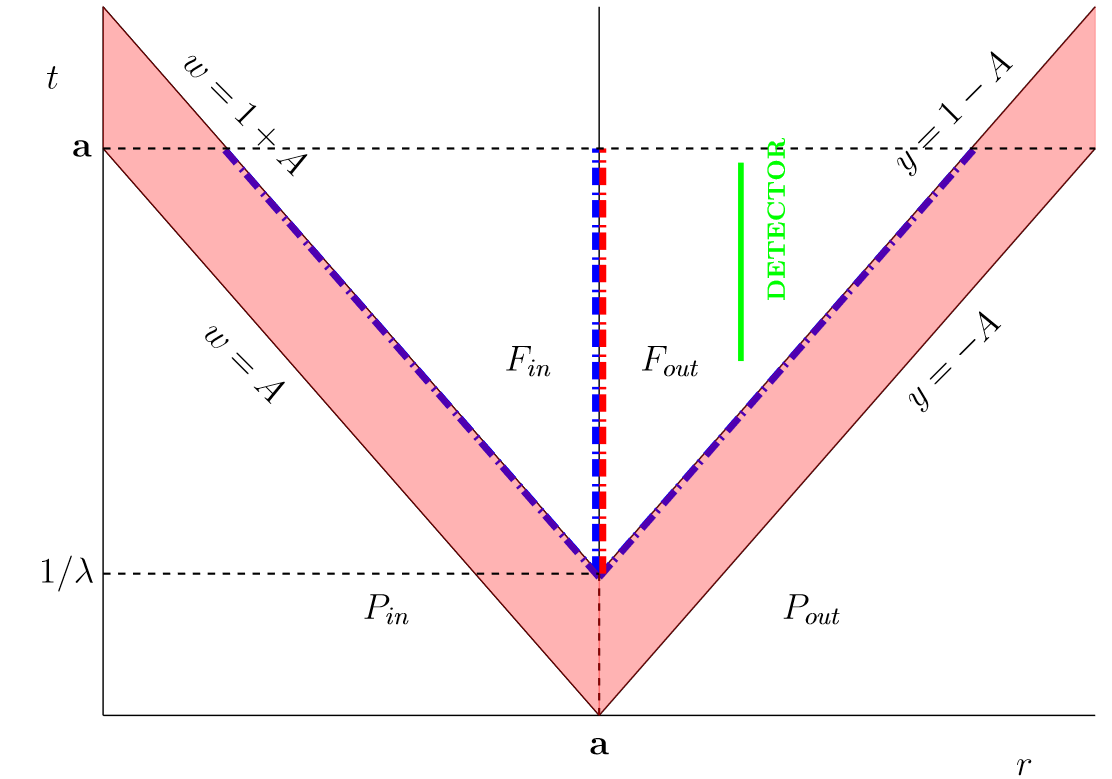}
\end{center}
\caption{
Spacetime diagram of the evolving boundary conditions.
The interpolation between $\theta=\pi/2$ and $\theta=0$ at $r=a$ 
occurs over $0<t<\lambda^{-1}$, and the null cones of the 
events where the boundary condition changes  
fill the regions $(-A<y<1-A,\,r>a)$ and $(A<w<1+A,\,r<a)$ in the spacetime. 
The triangular regions marked $P_{\rm out}$, $P_{\rm in}$, $F_{\rm out}$ and $F_{\rm in}$ indicate what we will refer to as the early time ($P$) and late time regions ($F$), outside and inside of the shell. 
The figure shows the case $a>\lambda^{-1}$. The blue dash-dotted line indicates that the energy density is positively divergent, while red dashed line shows where the energy density is negatively divergent.
\label{trplane-fig}}
\end{figure}

Rearranging (\ref{bc3}) we obtain first order differential equations for the derivatives $\partial_t E_k(t,a)$ and $\partial_t G_k(t,a)$. Introducing the additional dimensionless variables $K=k/\lambda$ and $A= a \lambda$, $y=\lambda u$ and $w=\lambda v$,  and defining the functions $R_K(y) = E_{k}(u)$ and $S_K(w) = G_{k}(v)$ we obtain
\bea
\label{Rresult}
R_K(y) && =\left\{
\begin{array}{ll}
 - e^{-i K y} & y<-A\\
 - e^{-i K y} + \frac{2i\sin(A K)}{B(y+A)} \int_0^{y+A} e^{-i K z} B^\prime (z)\,dz ~~~~& y\in(-A,1-A)\\
 - e^{-i K(y+2A)} & y>1-A
 \end{array} \right.
\eea
\bea
\label{Sresult}
S_K(w) && = \left\{
\begin{array}{ll}
 e^{-i K w} & w<A\\
  e^{-i K w} +  \frac{2i\sin(A K)}{B(w-A)} \int_0^{w-A} e^{-i K z} B^\prime (z)\,dz ~~~~ & w\in(A,1+A)\,. \\
  e^{-i K(w-2A)} ~~~~ & w > 1+A
\end{array} \right .
\eea
Substituting these expressions into (\ref{anz}) we obtain expressions for the mode functions $U_k(t,r)$ in the regions inside and outside of the shell. A spacetime diagram is shown in Fig. \ref{trplane-fig}. 
 
In order to write equations in a more compact form we sometimes use a shorthand notation in which functions that depend only on $y+A$ are written without their arguments, for example $B:=B(y+A)$. This notation is used throughout the appendices. 

The solutions in equations (\ref{Rresult}, \ref{Sresult}) have the following features:
\begin{enumerate}
\item The regions in figure \ref{trplane-fig} marked $P_{\rm out}$ and $P_{\rm in}$ correspond to the early time regions, relative to the wall creation. The top line of either (\ref{Rresult}) or (\ref{Sresult})  gives the mode functions in these regions, which are just those of the Minkowski vacuum: $U_k(t,r) = \frac{1}{\sqrt{4\pi k} }\,\big(e^{-i k v}-e^{-i k u}\big) = -\frac{i e^{-i k t}\sin(k r)}{\sqrt{\pi k} }$.
\item The regions in figure \ref{trplane-fig} marked $F_{\rm out}$ and $F_{\rm in}$ correspond to the late time regions, relative to the wall creation. The mode functions in these two regions are obtained from the bottom lines of (\ref{Rresult}, \ref{Sresult}) which give
\bea
U_k(t,r) &=& \frac{-ie^{-ik(t+a)}\sin[k(r-a)]}{\sqrt{\pi k}}\,, ~~~~ y>1-A ~\text{and}~ r>a~~[\text{region}~~F_{\rm out}]\,, \nonumber \\
U_k(t,r) &=& \frac{-ie^{-ik(t-a)}\sin[k(r-a)]}{\sqrt{\pi k}}\,, ~~~~ w>1+A ~\text{and}~ r<a~~[\text{region}~~F_{\rm in}] \,. \nonumber
\eea
Comparing with the above expression we see that when $ka=2n\pi$, which means that wall formation occurs at a node of the original mode, the mode function in the late time region is the same as the original Minkowski mode. 
\item  The middle lines of (\ref{Rresult}) and (\ref{Sresult}) give, respectively, the expressions for the mode functions on the inward and outward light cones 
of the events where the boundary condition changes 
(at $r=a$ and $t\in(0,\lambda^{-1})$). 
\item For $t<a$ we have
$$
\lim_{r\to a_-} \big(G_k(v)-e^{-i k u}\big) = \lim_{r\to a_+}\big(e^{-i k v}+E_k(u)\big)
$$
and therefore the mode functions $U_k(t,r)$ are continuous at the location of the shell $r=a$.
\item It is easy to see that $R_K(y)$ is smooth for $y<1-A$ and in Appendix \ref{smooth} we show that it is also smooth at $y=1-A$. Similarly $S_K(w)$ is smooth for $w\le 1+A$.
\item The mode functions are normalized so that 
$$
(\phi_k,\phi_{k^\prime}) = i \int_0^\infty dr\,r^2 \int_{S^2} d\Omega\;\big(\overline{\phi_k}\partial_t \phi_{k^\prime} - (\partial_t \overline{\phi_k})\phi_{k^\prime} \big) = \delta(k-k^\prime)
$$
for any constant time hypersurface with $t<\lambda^{-1}$.
\item The late time region inside the shell ($w>1+A$ and $r<a$ which is denoted region $F_{\rm in}$ in figure \ref{trplane-fig}) is not part of our calculation. 
One problem is that this part of the spacetime diagram would be influenced by waves reflected at the origin, so that we do not expect the ansatz (\ref{anz}) to be satisfied in this region. 
Physically, once the shell is fully formed, an infinite potential separates the regions inside and outside, which means there is no flow of probability between them.
\end{enumerate}
Useful alternate expressions for the functions $R_K(y)$ and $S_K(w)$ are obtained by integrating the middle lines in (\ref{Rresult}) and (\ref{Sresult}) by parts: 
\bea
\label{Ralternate}
&& R_K(y) = -e^{-i K(y+2A)}-\frac{2i\sin (AK)}{B(y+A)} - \frac{2 K\sin(A K)}{B(y+A)} \int^{y+A}_0 e^{-i K z} B(z)\,dz \,,\\
\label{Salternate}
&&S_K(w) = e^{-i K(w-2A)}-\frac{2i\sin(AK)}{B(w-A)}  - \frac{2 K\sin(A K)}{B(w-A)} \int^{w-A}_{0} e^{-i K z} B(z)\,dz\,.
\eea
We note that the mode functions in the future regions $F_{in}$ and $F_{out}$, from the third lines of (\ref{Rresult}) and (\ref{Sresult}), are $\lambda$ independent. As we will see in the section \ref{sec:detector}, this means that the detector response in the future regions does not depend on $\lambda$, which is the parameter that controls how fast the shell evolves at $r=a$.

\section{Total energy}
\label{section-energy}

The expression for the energy density outside of the shell in terms of the mode functions $R_K(y)$ has exactly the same form as in Ref. \cite{Zhou}. We summarize the calculation and give the result below. The renormalized energy density of the quantised field in the state $|0\rangle$ is obtained by point-splitting the field operators, taking the expectation value in $|0\rangle$, subtracting the corresponding expectation value in the Minkowski vacuum $|0\rangle_M$, and taking the coincidence limit. This gives
\bea
\langle T_{00}\rangle &&= \langle 0|T_{00}|0\rangle_{\rm ren} \nonumber \\
&& = \lim_{\substack{u_1,u_2\rightarrow u\\v_1,v_2\rightarrow v}}(\partial_{u_1}\partial_{u_2}+\partial_{v_1}\partial_{v_2})\Big[\langle 0|\phi(1)\phi(2)|0\rangle-\langle 0_M|\phi(1)\phi(2)|0_M\rangle\Big]\nonumber\\
&& =\frac{1}{4\pi}\bigg(\frac{\langle(\partial_u f)^2\rangle}{r^2}+\frac{\langle f(\partial_u f-\partial_v f)\rangle+c.c}{2r^3}+\frac{\langle f^2\rangle}{2r^4}\bigg)\,.
\eea
Using Eq. (\ref{eq:f-op}) and the first line of (\ref{anz}) we obtain in the outside region ($r>a$)
\begin{equation}\label{stress-energy-concreteform}
\langle T_{00}\rangle=\frac{\lambda^2}{16\pi^2 r^2}\int_{0}^{\infty}\frac{dK}{K}[|R_K'(y)|^2-K^2]-\frac{1}{32\pi^2r^2}\frac{\partial}{\partial r}\Big(\frac{\mathcal{G}_{\rm out}(t,r)}{r}\Big)
\end{equation}
where $\mathcal{G}_{\rm out}$ is defined as
\begin{equation}\label{G function}
\mathcal{G}_{\rm out}=\int_{0}^{\infty}\frac{dK}{K}\Big[|e^{-iKw}+R_K(y)|^2-|e^{-iKw}-e^{-iKy}|^2\Big]\,.
\end{equation}
Using the second line of (\ref{anz}) gives for the inside region ($r<a$)
\begin{equation}\label{stress-energy-inside-concreteform}
\langle T_{00}\rangle=\frac{\lambda^2}{16\pi^2 r^2}\int_{0}^{\infty}\frac{dK}{K}[|S_K'(w)|^2-K^2]-\frac{1}{32\pi^2r^2}\frac{\partial}{\partial r}\Big(\frac{\mathcal{G}_{\rm in}(t,r)}{r}\Big)
\end{equation}
with the definition
\begin{equation}\label{G function-inside}
\mathcal{G}_{\rm in}=\int_{0}^{\infty}\frac{dK}{K}\Big[|S_K(w)-e^{-iKy}|^2-|e^{-iKw}-e^{-iKy}|^2\Big]\,.
\end{equation}

In the early time regions, which are denoted $P_{\rm out}$ and $P_{\rm in}$ in figure \ref{trplane-fig}, $\langle T_{00}\rangle$ vanishes by construction. 
In the late time regions $F_{\rm out}$ and $F_{\rm in}$, the first terms in (\ref{stress-energy-concreteform}) and (\ref{stress-energy-inside-concreteform}) vanish, as can be seen from (\ref{Rresult}, \ref{Sresult}).
The functions ${\cal G}_{\rm out}$ and ${\cal G}_{\rm in}$ are easily calculated.
We obtain
\bea
\label{after-T00}
&& {\cal G}_{\rm out\,future} = {\cal G}_{\rm in\,future} = \ln\left[(r-a)^2\right] - \ln(r^2) \\[2mm]
&& \langle T_{00}\rangle_{\rm future} =\frac{1}{16\pi^2 r^4}\Big(\ln |r-a| -\ln (r)\Big)-\frac{1}{16\pi^2 r^3}\Big(
\frac{1}{r-a}-\frac{1}{r}\Big)\,.
\label{after-T00-2}
\eea
This result shows that the energy density goes discontinuously from positive infinity just inside the shell to negative infinity energy  just outside the shell. However, the spatial integral of the energy density along a constant time hypersurface which crosses $r = a$ produces a finite energy, if interpreted at $r=a$ in the principal value sense.
In Appendix \ref{outside-region} we consider the intermediate outside region $r>a$ and $-A < y < 1-A$ and consider the continuity of the energy density as $y+A\to 1_-$.
We show that the energy density is finite everywhere except to the immediate past of the light cone of the point where boundary condition finishes changing, where it is positively divergent. 

\section{Detector Response}\label{sec:detector}

In this section we consider an inertial 
Unruh-DeWitt (UDW) detector \cite{Unruh:1976db,DeWitt:1979} that is linearly coupled to the quantum field and 
at a fixed spatial location. 
Using first-order perturbation theory, the probability that the 
detector undergoes a transition from a state with energy $0$ 
to a state with energy $\omega$ is proportional to the response function 
\cite{Unruh:1976db,DeWitt:1979,birrell-davies,wald-smallbook} 
\begin{align}
\mathcal{F}(\omega) =
\int_{-\infty}^{\infty}dt_1\int_{-\infty}^{\infty}dt_2 
\, e^{-i\omega(t_1-t_2)}
\, \chi(t_1)\chi(t_2) \, \mathcal{W}(t_1, t_2)
\,,
\label{eq:F-genformula}
\end{align}
where the smooth real-valued switching function $\chi$ specifies how the 
detector's interaction with the field is turned on and off, and 
$\mathcal{W}$ is the pull-back of the field's Wightman function 
to the detector's worldline. 
We consider a detector located outside the spherical shell ($r>a$) and operating only in the future region $y>1-A$. 
The response function is 
\begin{align}
\Delta \mathcal{F}(\omega) =
\int_{-\infty}^{\infty}dt_1\int_{-\infty}^{\infty}dt_2 
\, e^{-i\omega(t_1-t_2)}
\, \chi(t_1)\chi(t_2) \, \Delta \mathcal{W}(t_1, t_2)
\,,
\label{eq:Delta-F-gen}
\end{align}
where 
\begin{align}
\Delta \mathcal{W}(t_1, t_2)
&= \frac{1}{4\pi r^2}
\int_0^\infty 
\Bigl(
U_k(t_1-r, t_1+r) \overline{U_k(t_2-r, t_2+r)} 
\notag
\\
&\hspace{14ex}
- 
U^M_k(t_1-r, t_1+r) \overline{U^M_k(t_2-r, t_2+r)} 
\, 
\Bigr) 
\, dk 
\,.
\label{eq:Delta-W-def}
\end{align}
The functions $U_k$ and $\overline{U_k}$ are obtained from the first line in (\ref{anz}) with $E_k(u) = R_K(y)$ and $R_K(y)$ given by the last line of (\ref{Rresult});
$U^M_k$ and $\overline{U^M_k}$ are obtained in the same way except using the first line of (\ref{Rresult}). Substituting these expressions and collecting terms we find
\bea
16\pi^2 r^2 \Delta \mathcal{W}(t_1,t_2)=\int_0^\infty \frac{dk}{k}\left(
e^{i k \left(2 r-t_1+t_2\right)}
+e^{-i k \left(2 r+t_1-t_2\right)}
-e^{i k \left(2 a-2 r-t_1+t_2\right)}
-e^{-i k \left(2 a-2 r+t_1-t_2\right)}\right)\,.\nonumber\\
\eea
Doing the $k$ integral we get
\bea
\label{resultdet2}
16\pi^2 r^2 \Delta \mathcal{W}(t_1,t_2) &=& \ln\left[
\frac{|4(a-r)^2-(t_1-t_2)^2|}{|4r^2-(t_1-t_2)^2|}
\right] \\[4mm]
&+& i\pi\, \text{sgn}(t_1-t_2) \theta \left(r-\left|t_1-t_2\right|/2\right) \theta \left(a-r+ \left|t_1-t_2\right| /2\right)\,.\nonumber
\eea   
The change in the Wightman function is $\lambda$ independent, and therefore one does not obtain a divergent response in the limit $\lambda\to0$, which would correspond to instantaneous wall creation. There is a divergence when $|t_1-t_2|=2r$, but this divergence is only logarithmic.

\section{Conclusion}
\label{summary}

We have extended previous work on firewall creation via time dependent boundary conditions by considering  the smooth and sharp creation of a fixed radius  shell in (3+1)-dimensional Minkowski space. This was implemented by introducing at $r=a$ a one parameter family of self-adjoint extensions of the three dimensional Laplacian operator that interpolate between 
the condition that the 
values and derivatives on the two
sides of $r=a$ coincide 
for $t\le0$ (no wall) and the two-sided Dirichlet boundary condition for $t \ge 1/\lambda$ (fully-developed wall).
Wall creation produces  null pulses of energy on either side of the shell, one ingoing and the other outgoing. The process is significantly more divergent than in (1+1)-dimensions \cite{brown}, but less divergent than for point-like source creation in (3+1)-dimensions \cite{Zhou}.  
In the present case, the energy density $\langle T_{00}\rangle$ diverges 
to positive infinity in the outgoing energy pulse, just outside the light cone of the fully-formed wall at $t=1/\lambda$. 
Unlike in
the 3+1 point source creation case, there is no persistent memory cloud of energy. As in the 1+1 dimensional wall case, the response of the detector 
in the post-shell region
is independent of the time scale $1/\lambda$ and finite, casting doubt once again on the viability of wall creation as a possible mechanism for breaking entanglement at the event horizon of an evaporating black hole.

\medskip

\section*{Acknowledgements}
This work was funded in part by the Natural Sciences and Engineering Research Council of Canada (MEC and GK) and by the Science and Technology Facilities Council (JL, Theory Consolidated Grant ST/P000703/1). For hospitality, GK thanks the University of Nottingham, and JL thanks the University of Winnipeg and the Winnipeg Institute for Theoretical Physics.

\appendix

\section{Scalar Laplacian on $\BbbR^3$ with a spherical shell\label{app:laplaceoperator}}

Defining the short hand notation $\fint^\infty_0dr  = \int_0^{a^-} + \int_{a^+}^\infty$ and scaling $g=f/r$,
we can map the 3 dimensional Laplacian and the $L_2$ inner product on a positive half line with a spherical shell at $r=a$ to  
\bea
\nabla^2
&&\Rightarrow\nabla^2 = \partial_r^2 +\frac{1}{r^2} \nabla^2_{S^2}\,\,,\,\,
(g_1,g_2)
\Rightarrow 
(f_1,f_2)= \int_{S^2} d\Omega\;\fint^\infty_0 dr\, \overline{f_1} f_2\,.
\eea
We consider only the spherically symmetric case (see equation (\ref{eq:phi-scaledto-f})), 
and Hermiticity of $\nabla^2$ requires
\bea
\label{herm1}
 (f_1,\nabla^2f_2)-(\nabla^2f_1,f_2)= \fint^\infty_0dr\, \partial_r \big[\overline{f_1}(\partial_r f_2) - (\partial_r\overline{f_1})f_2\big]
 = 0\,.
\eea
Taking $f_1=f_2\equiv f$, the square bracket $\big[\overline{f_1}(\partial_r f_2) - (\partial_r\overline{f_1})f_2\big]$ can be written
\bea
S(t,x)  = \frac{1}{(2 i L)}\big(|L \partial_r f - i f|^2 -|L \partial_r f + i f|^2\big)\,.
\eea
Following the method in \cite{bonneau}, we find that the boundary conditions are given by
\begin{equation}
\label{introU}
\left( \begin{array}{c}
Lf'_{+} -if_{+}\\
Lf'_{-} +if_{-}\\
Lf'_{0} -if_{0}
\end{array} \right)
=U
\left( \begin{array}{c}
Lf'_{+} +if_{+}\\
Lf'_{-} -if_{-}\\
Lf'_{0} +if_{0}
\end{array} \right) 
\end{equation}
where we have used the shorthand notation $f(t,0)=f_0$, $f(t,a^-)=f_-$ and $f(t,a^+)=f_+$, 
and the choice of a unitary 3 $\times$ 3 matrix $U$ specifies the boundary condition.
Generally, the matrix $U$ would be decomposed by Gell-man matrices.  
We choose an expression of $U$ that 
ensures no flow of probability through the origin and depends on one free parameter which is a time dependent function chosen to model the formation of the wall at $r=a$.  We use
\begin{equation}
U=\left( \begin{array}{ccc}
e^{-i\theta(t)}\left(\begin{array}{cc} 
\cos\theta(t) & i\sin\theta(t)\\
i\sin\theta(t) & \cos\theta(t)
\end{array}\right)\\
 & & 1
\end{array} \right)
\end{equation}
and equation (\ref{introU}) gives the conditions
\bea
f(0)&&=0 \\
f_-&&=f_+ \\
\frac{2}{L}\cot\theta(t)&&= \frac{f'_{+}}{f_{+}}-\frac{f'_{-}}{f_{-}}
\eea
We note that the continuity of the wave function at $r=a$ is a result of our choice of $U(\theta)$, and does not have to be imposed as an extra condition. 

\section{Mode function regularity}\label{smooth}

In this appendix we show that the mode function $R_K(y)$ is $C^{26}$ at $y=1-A$. 
 We follow closely Appendix B of \cite{Zhou}.

\subsection{Smoothness of $1/B(y)$}
\label{Bsmooth}

First we show that $1/B(y)$ and all its derivatives go to zero as $y\to 1_-$. 
We write $g(y) := \tan\bigl(h(y)\bigr)$ and from equation (\ref{eq:theta-scaled}) we see that $g(y)>0$, and $g(y)$ and all its derivatives approach $0$ as $y\to1_-$. 

Using 
\eqref{eq:B-sol} we obtain 
\bea
B(y) &&= \exp\left(\int_{0}^{y} \frac{dz}{g(z)} \right)\\
B^\prime(y) &&=B(y)/g(y)\label{eq:B-ito-g}\,.
\eea
Defining $B_{\rm in}(y) = 1/B(y)$, we have $B^\prime_{\rm in}(y)= -B_{\rm in}(y)/g(y)$. For $n \in \BbbN$, induction gives
\begin{subequations}
\begin{align}
B_{\rm in}^{(n)}(y) &= (-1)^n
P_n(y) f_n(y)
\,, 
\\
f_n(y) &= 
\frac{B_{\rm in}(y)}{\bigl(g(y)\bigr)^n} 
\,,
\label{eq:fn-def}
\end{align}
\end{subequations}
where each $P_n$ is a polynomial in $g$ and its derivatives and $P_n(y) \to 1$ as $y\to 1_-$. 
We show below that $f_n(y) \to 0_+$ as $y\to1_-$, and therefore $B^{(n)}_{\rm in}(y) \to 0$.
From \eqref{eq:fn-def}
we have 
\begin{align}
\ln\bigl(f_n(y)\bigr) =
-\left(\int_{0}^{y} \frac{dz}{g(z)} \right) 
\left(1 + \frac{n \ln\bigl(g(y)\bigr)}{\displaystyle\int_{0}^{y} 
\frac{dz}{g(z)}} \right) 
\,.
\label{eq:lnf-formula}
\end{align}
As $y\to1_-$, the first parentheses in \eqref{eq:lnf-formula} tend to~$\infty$, 
while the second parentheses tend to $1$ by l'H\^opital. 
Hence $\ln\bigl(f_n(y)\bigr) \to -\infty$ 
as $y\to1_-$ and therefore $f_n(y) \to 0_+$ as $y\to1_-$. 

\subsection{Differentiability of $R_K(y)$}\label{Rsmooth}
We make the definitions 
\bea
&& J_K(y) = \int^y_0 {B}(z) \, e^{-iK z} \, dz \\[2mm]
\label{defF}
&& {\cal F}_K(y) = J_K(y)/B(y) \\[2mm]
\label{defH}
&& {\cal H}_K(y) = J_K(y)/\left(g(y)\,B(y)\right)\,,
\eea
and rewrite the mode function (\ref{Ralternate}) as
\begin{subequations}
\label{RSalternate2}
\begin{align}
& R_K(y) = -e^{-i K(2A+y)} -\frac{2i\sin(AK)}{B} - 2 K\sin(A K) {\cal F}_K(y+A) \,.
 \end{align}
\end{subequations} 
We will show that $R_K(y)$ is $C^{26}$ at $y+A\to 1_-$ by showing that ${\cal F}_K(y)$ is $C^{26}$ as $y \to 1_-$.  We outline the steps below. 
\begin{enumerate}
\item Calculate derivatives ${\cal F}_K^{(n)}(y)$ using (\ref{eq:B-ito-g}) to remove factors of $B^\prime(y)$, so that the result does not contain derivatives of $B(y)$.
\item Define ${\rm den}(y) = B(y)\big(g(y)\big)^n$ and ${\rm num}(y) = {\cal F}_K^{(n)}(y) \,{\rm den}(y)$.
In order to use l'H\^opital $n$ times, calculate ${\rm num}^{(n)}(y)$ and ${\rm den}^{(n)}(y)$, again using (\ref{eq:B-ito-g}) after taking each derivative. 
\item Define ${\bf num}^{(n)}(y) = {\rm num}^{(n)}(y)/B(y)$ and ${\bf den}^{(n)}(y) = {\rm den}^{(n)}(y)/B(y)$ and replace all remaining factors of $J_K(y)$ by ${\cal F}_K(y)\,B(y)$. 
The limit of the $n$-th derivative of ${\cal F}(y)$ as $y\to 1_-$ now has the form
\bea
\label{resultF}
\lim_{y\to 1_-} {\cal F}^{(n)}(y) \xrightarrow{\text{l'H\^opital} } \lim_{y\to 1_-} \frac{{\bf num}^{(n)}(y)}{{\bf den}^{(n)}(y)}\,.
\eea
\end{enumerate}
Using computer algebra we have verified up to order $n=26$ that the quantities ${\bf num}^{(n)}(y)$ and ${\bf den}^{(n)}(y)$ have the form
$$
{\bf num}^{(n)}(y) = P(y)\,,~~~
{\bf den}^{(n)}(y) = 1+Q(y)
$$
where $P(y)$ and $Q(y)$ are polynomials of $g(y)$ and its derivatives, ${\cal F}(y)$, and factors $e^{-i K y}$, and each term contains at least one power of 
${\cal F}(y)$ or $g^{(j)}(y)\,~j\in(0,n)$. This means that both $P(y)$ and $Q(y)$ go to 0 as $y\to 1_-$. We therefore have from (\ref{resultF}) that ${\cal F}(y)$ is $C^{26}$ at $y=1$. We stopped the calculation at $n=26$ because of limitations of computing time and memory. If the proof extends to $n\in \BbbN$, we would have that $R_K(y)$ is smooth at $y=1$.

\section{Lemma and conditions on $g$\label{sec:g-conditions}}

\begin{lem}\label{lemma-a}$\phantom{xxx}$ 
	\begin{enumerate}
		\item[(i)]
		For a complex-valued function $f(y,z)$ that is bounded for $y\in[0,1]$ and $z\in[0,1+A]$, we have 
		\begin{equation}\label{eq:lemma-a}
		\lim\limits_{y+A\rightarrow 1_-}\frac{1}{B(y+A)} \int_{0}^{y+A}B(z)f(y,z) \, dz = 0\,.
		\end{equation}
		\item[(ii)]
		If in addition $\partial_y f(y,z)$ is bounded for $y\in[0,1]$ and $z\in[0,1+A]$, we have 
		\begin{equation}\label{eq:lemma-b}
		\lim\limits_{y+A\rightarrow 1_-} \partial_y \left( \frac{1}{B(y+A)} \int_{0}^{y+A}B(z)f(y,z) \, dz \right) = 0\,.
		\end{equation}
	\end{enumerate}
\end{lem}
{\it Proof}. 

(i) By boundedness of~$f$, 
there is a positive constant $C$ such that $|f(y,z)| \le C$. For $0 < y+A<1$, using the triangle inequality and the positivity of~$B$, we have 
\begin{align}
\left| \frac{1}{B(y+A)} \int_{0}^{y+A}B(z)f(y,z) \, dz \right | 
\le \frac{C}{B(y+A)} \int_{0}^{y+A}B(z) \, dz \,.
\label{lemma-aa1}
\end{align}
When $y+A \to 1$, the rightmost expression in \eqref{lemma-aa1} 
goes to zero, using l'H\^opital and~\eqref{eq:B-ito-g}. 

(ii) For $0 < y+A<1$, we expand out the derivative 
in \eqref{eq:lemma-b} using \eqref{eq:B-ito-g} and make some cancellation. After repeatedly applying l'H\^opital as $y+A \to 1$ and using lemma \eqref{eq:lemma-a}, the proof is done. 
$\blacksquare$

\begin{cond}\label{g-condition}\hspace*{1cm}\\[2mm]
We introduce the technical assumption that $g'''(y) < 0$ for $0<y<1$. 
It follows that $g''(y) > 0$ and $g'(y) < 0$ for $0<y<1$. 
For $0 < z \le y+A < 1$, this implies 
\begin{align}
&g'(z)  \leq \frac{g(y+A)-g(z)}{y+A -z} \leq g'(y+A)\,, 
\label{eq:g condition} 
\\
&g''(y+A)  \leq \frac{g'(y+A)-g'(z)}{y+A -z} \leq g''(z)\,, 
\label{eq:g' condition}
\end{align}
where the quotients are understood at $z= y+A$ in the limiting sense. 
\eqref{eq:g condition} can be 
verified by writing the numerator as the integral of 
$g'$ and using the monotonicity of~$g'$, 
and \eqref{eq:g' condition} can be verified similarly be 
writing the numerator as the integral of~$g''$. 
We will derive three consequences which are used in sections \ref{Fsection} and \ref{Ftilde-section}.
\end{cond}

\begin{enumerate}[i.]
	\item \textbf{First consequence: $\lim_{y+A\to 1_-}\frac{g'(y+A)}{g(y+A)} = -\infty $.}
	
	For $0 < y+A < 1$, using the monotonicity of~$g'$, we have 
	\begin{align}
	g(y+A) = -\int^1_{y+A} g^\prime(z) dz \le -g^\prime(y+A)\int^1_{y+A}  dz 
	= - (1 - y-A) g^\prime(y+A) 
	\,.
	\end{align}
	Hence 
	$g'(y+A)/g(y+A) \le - 1/(1-y-A)$, which implies $\lim_{y+A\to 1_-}\frac{g'(y+A)}{g(y+A)} = -\infty $.
	\item  \textbf{Second consequence: $\lim_{y+A\to 1_-} {\cal C}(y) = 0$}
	
	For $0 < y+A < 1$, we define
	\bea
	{\cal C}(y) && =\frac{1}{B(y+A)}\int_0^{y+A} dz \cos(y+A-z) B^\prime(z)\;\frac{g(y+A)-g(z)}{y+A-z} \,,\label{Cdef}\\
	{\cal C}_-(y) &&= 
	\frac{1}{B(y+A)}\int_0^{\kappa} dz  \left[\frac{g(y+A) B^\prime(z)}{y+A-z} - \frac{B(z)}{y+A-z} \right] \,, \label{eq:Ctildeminus-def}\\
	{\cal C}_+(y) &&= \frac{1}{B(y+A)}\int^{y+A}_{\kappa} dz  B^\prime(z)\;\frac{g(y+A)-g(z)}{y+A-z}\,,\\
	\tilde{{\cal C}}_-(y) &&=   -\frac{1}{B(y+A)}\int_0^{\kappa} dz \frac{B(z)}{y+A-z}\,,  \\
	\tilde{{\cal C}}_+(y) &&=   \frac{1}{B(y+A)}\int_{\kappa}^{y+A} dz B^\prime(z)g^\prime(z)\,,
	\eea
	where $0 < \cos(y+A-z) \le 1$, $B^\prime(z)>0$ and $g^\prime(z)<0$ for $0<z<y+A$, and constant 
	$\kappa$ with $0<\kappa < y+A < 1$. 
	We have $\tilde{{\cal C}}_-(y) < {\cal C}_-(y)$ and $\tilde{{\cal C}}_+(y) \le {\cal C}_+(y)$ using \eqref{eq:g condition}. Therefore
	\bea
	\tilde{{\cal C}}_+(y)+\tilde{{\cal C}}_-(y)<{\cal C}_+(y)+{\cal C}_-(y)<{\cal C}(y)<0  
	\label{eq:C3C2C-ineq}
	\eea
	As $y+A \to 1_-$, both integral $\tilde{{\cal C}}_-(y)$ and integral $\tilde{{\cal C}}_+(y)$ tend to $0$. 
	From \eqref{eq:C3C2C-ineq} it then follows that $\lim_{y+A\to 1_-} {\cal C}(y) = 0$.
	\item \textbf{Third consequence: $\lim_{y+A\to 1_-}{\cal D}(y) = 0$}
	
	We define
	\bea
	{\cal D}(y) &&= \frac{1}{B}\int_\kappa^{y+A} dz\,B'(z)\frac{\cos (\alpha )}{\alpha} \left(\frac{g(A+y)-g(z)}{\alpha} -g'(A+y)\right)\label{Ddef}\,,\\
	{\cal D}_2(y)&& = \frac{1}{B}\int_\kappa^{y+A} dz\,B'(z)\frac{1}{\alpha} \left(\frac{g(A+y)-g(z)}{\alpha} -g'(A+y)\right)\,,\\
	{\cal D}_3(y)&& = \frac{1}{B}\int_\kappa^{y+A} dz\,B'(z)\frac{1}{\alpha} \left(g'(z) -g'(A+y)\right)\,,
		\eea
	\bea
		{\cal D}_4(y)&& = -\frac{1}{B}\int_\kappa^{y+A} dz\,B'(z) g''(z) \,,
	\eea 
	where $0 < \cos(y + A - z) \le 1$ and $B'(z) > 0$ for $0 < z < y + A$. Using equation (\ref{eq:g condition}) and (D.2), there is
	\bea
	{\cal D}_4(y)\le{\cal D}_3(y)\le{\cal D}_2(y)<{\cal D}(y)<0\,.
	\eea
	Further use of l'H\^opital on ${\cal D}_4(y)$ we find $\lim_{y+A\to 1_-}{\cal D}_4(y)=0$, which means $\lim_{y+A\to 1_-}{\cal D}(y)=0$.
	\end{enumerate}

\section{Behaviour of $\langle T_{00}\rangle$ outside the shell in the intermediate region} 
\label{outside-region}

In this section we study the behaviour of the energy density outside the shell ($r>a$) in the intermediate region ($-A\le y \le 1-A$), which is given in Eq. (\ref{stress-energy-concreteform}) with the function ${\cal G_{\rm out}}$ defined in (\ref{G function}), and $R_K(y)$ given in the middle line of (\ref{Rresult}).
We rewrite these expressions as
\bea
\label{T00outside}
&& \langle T_{00}\rangle = \frac{\lambda^2}{16\pi^2 r^2}\big[F(y)+\tilde F(y)\big] \\[2mm]
&& F(y) = \int_{0}^{\infty} \frac{dK}{K} \; \big[|R'_K(y)|^2-K^2\big] \label{Fdef}\\[2mm]
&& \tilde F(y) = -\frac{1}{w-y}\left[\partial_w-\partial_y-\frac{2}{w-y}\right]{\cal G}_{\rm  out}\,,\label{Ftildedef}
\eea
where the notation suppresses the dependence of $F(y)$ and $\tilde F(y)$ on $w$. 
We comment that the denominators in (\ref{Ftildedef}) are produced by the factor $1/r$ in equation (\ref{stress-energy-concreteform}), and the action of the derivative $\partial_r$ on this factor.

\subsection{Preliminaries}
From (\ref{Ralternate}) we have for fixed $y\in(-A,1-A)$ the small $K$ estimates
\begin{align}
R_K(y)&=-1+O(K)\,,\\[2mm]
|R'_K(y)|^2&=O(K^2)\,,
\end{align}
and the large $K$ estimates
\begin{align}
|R_K(y)|^2& =1+\frac{2B'}{B}\frac{\sin(2AK)}{K}+O(K^{-2})\,,\\[2mm]
|R'_K(y)|^2&=K^2+\frac{2B'}{B}K\sin(2AK)+ O(K^{-1})\,.\label{2check}
\end{align}
Using these results it is straightforward to show that the integrals in equations (\ref{Fdef}) and (\ref{Ftildedef}) are well defined for $y\in(-A,1-A)$.

At $y=-A$ both integrands vanish. 
At $y=1-A$ the integrand in (\ref{Fdef}) vanishes, and a straightforward calculation gives
${\cal G}_{\rm out}\big|_{y=1-A} = 2\big[\ln(r-a)-\ln r\big]$ in agreement with (\ref{after-T00}). 
In the remainder of this appendix we study the continuity of $F(y)$ and $\tilde F(y)$ in equations (\ref{Fdef}) and (\ref{Ftildedef}) as $y+A\to 1_-$. 

\subsection{Term $F(y)$}\label{Fsection}
We divide the integral in (\ref{Fdef}) into two pieces defined as
\bea
F_-(y) & =&\int_{0}^{1}\frac{dK}{K} \; \big[|R'_K|^2-K^2\big]\label{F_-}\,,\\
F_+(y) & =&\int_{1}^{\infty}\frac{dK}{K} \; \big[|R'_K|^2-K^2\big]\,.\label{F_+} \label{Fplus}
\eea
To show that $F_-(y)$ is finite, we use the form of $R_K(y)$ in equation (\ref{Ralternate}). 
Differentiating and substituting into (\ref{F_-}), we get an expression with ${\cal H}_K$ as defined in Eq.(\ref{defH}) and $f_i :=f_i(y+A)$ as defined in Eq.(\ref{eq:fn-def}).
It is then straightforward to show with l'H\^opital that $\lim_{y+A \to 1_-}F_-(y)$ exists and is finite.\\

Now we study the second term $F_+(y)$. Our strategy is as follows:
\begin{enumerate}
\item Integrate by parts in $z$, taking the anti-derivative of the factor $e^{\pm i K z}$, until we have enough powers of $1/K$ so that the $K$-integral is convergent. 
\item Switch the order of the $z$ and $K$ integrals and do the $K$ integral. 
\item Integrate by parts again so that derivatives are removed from factors $B(z)$. 
\item Analyse the behaviour of the remaining integrals. 
\end{enumerate}
Defining 
\bea\label{Vdef}
V_K = \frac{1}{B}\int_0^{y+A} dz\,B^{\prime\prime}(z) e^{-i K z}\,,
\eea
the result of step (1) is 
\bea
F_+(y) = \int_1^\infty dK \bigg[ 
&& \frac{4 V_K \overline{V_K} \sin ^2(A K) }{K^3 g^2} 
-  \frac{4 \sin ^2(A K)  \left(V_K e^{i K (A+y)}+\overline{V_K} e^{-i K
   (A+y)}\right)}{K^3 g^3}\nonumber\\
+ && \frac{4 i \sin ^2(A K) \left(V_K e^{i K (A+y)}-\overline{V_K} e^{-i K
   (A+y)}\right)}{K^2 g^2} \nonumber\\
+ && \frac{2 i \sin (A K)  \left(V_K e^{i K y}-\overline{V_K} e^{-i K y}\right)}{K
   g} \nonumber\\
+ && \frac{4 \sin ^2(A K) }{K^3 g^4} + \frac{2 \sin (2 A K) }{g} 
 \bigg]\,.\label{parts1}
\eea
The integral over large $K$ has terms of the form $\int^\infty_1 dK\sin (2AK)$ which is regularized with a factor $e^{-\delta K}$ to obtain a finite result in the limit $\delta\rightarrow 0$.
The other $K$ integrals can be done using
\bea\label{K-integrals}
&& \int_1^\infty \frac{dK}{K} \big(e^{i K \alpha} + e^{-i K \alpha} \big) = -2 {\rm Ci}(\alpha) \\
&& \int_1^\infty \frac{dK}{K^2} \big(e^{i K \alpha} + e^{-i K \alpha} \big) = 2 \cos(\alpha)+2\alpha {\rm Si}(\alpha) - \pi \alpha \\
&& \int_1^\infty \frac{dK}{K^3} \big(e^{i K \alpha} + e^{-i K \alpha} \big) = \cos(\alpha) - \alpha \sin(\alpha) + \alpha^2 {\rm Ci}(\alpha) 
\eea
where $\alpha>0$ and ${\rm Ci}$ and ${\rm Si}$ are the 
cosine and sine integrals in the notation of~\cite{dlmf}.

First, for terms that include $V_K$ as defined in \eqref{Vdef}, we interchange the integration by the absolute convergence of multiple integral, and we evaluate the integration over $K$ to reach an expression containing elementary functions and expressions like \eqref{K-integrals}. 
Many terms would contain factor $B'$ or $B''$, but we simply integrate by part to reduce them to combinations involving \eqref{eq:lemma-a} and \eqref{eq:lemma-b}. 

Next we introduce the parameters $\alpha := A+y-z$, $\beta := A-y+z$, $\gamma := 3A+y-z$, which will be used in 2nd, 3rd and 4th terms of (\ref{parts1}) for which we have $z\in(0,y+A)$; and 
$\rho := 2 A - x + z$, $\sigma:= 2 A + x - z$ and $\tau := x - z$ 
for the 1st term of (\ref{parts1}) where $x\in(0,y+A)$ and $z\in(0,x)$.
Using our original assumption $A>1$ these parameters are all positive over the full range of the corresponding integrals. We therefore find out that single integrals not having factors $\text{Ci}(\alpha)$ and double integrals not having $\text{Ci}(\tau)$ can be easily proven to be bounded when $y+A\to 1_-$.

Third, we consider the remaining contributions to (\ref{Fplus}) which contain factors Ci$(\alpha)$ and Ci$(\tau)$. These are the difficult terms and we label them $I_{hard}$.
The terms containing Ci$(\tau)$ are double integrals of variable $x$ and $z$, we handle them by first integrating by part in $z$ then $x$ and then  integrating by part in $x$ then $z$. This two different ways of integration by parts generate two equivalent expressions of the double integral. A lot of terms cancel out when we represent the double integral using its two equivalent expressions, further use of Lemma \ref{lemma-a}, $\lim_{y+A\to 1_-}{\cal C}^+(y)=0$ and $\lim_{y+A\to 1_-}{\cal D}^+(y)=0$ in appendix \ref{g-condition} eventually reduce the result to 
\bea
\label{hard-res}
\lim_{y+A\to 1_-} I_{\rm hard} = -\lim_{y+A\to 1_-}\frac{g'}{g^2}\,.
\eea
Therefore $F(y)$ is divergent as $y+A\to 1_-$.

\subsection{The term $\tilde F(y)$}
\label{Ftilde-section}

The integral in (\ref{Ftildedef}) can be divided into two pieces
\bea
\label{Gsmall}
\mathcal{G}^-_{\rm out} &=& \int_{0}^{1}\frac{dK}{K}\Big[|e^{-iKw}+R_K(y)|^2-|e^{-iKw}-e^{-iKy}|^2\Big]\,, \\[4mm]
\label{Gbig}
\mathcal{G}^+_{\rm out} &=& \int_{1}^{\infty}\frac{dK}{K}\Big[|e^{-iKw}+R_K(y)|^2-|e^{-iKw}-e^{-iKy}|^2\Big]\,.
\eea
We use two different forms for $R_K(y)$:
in $\mathcal{G}^-_{\rm out} $ we use equation (\ref{Ralternate}), and in $\mathcal{G}^+_{\rm out}$  we use equation (\ref{Rresult}). 

First we look at the easy piece (\ref{Gsmall}). We define the factors
\bea
&& l_1(z) =\frac{\cos(z)-1}{z}\,,\,l_{2}(z) =\frac{\cos (z)-1}{z^2}+\frac{\sin (z)}{z}\,,
\eea
and in addition to the definitions $\alpha\,,\beta\,,\gamma\,,\rho\,,\sigma\,,\tau$, we use $\mu:=2A-z$, $\nu:=2A+z$, $\xi := A+w-z$ and $\chi := w-z-A$. It is straightforward to show that the parameters $\mu$, $\nu$, $\xi$, and $\chi$ are all non-negative for $A>1$, $z\in(0,y+A)$ and $y\in(-A,1-A)$.
We also note that $l_1(z)$ and $l_2(z)$ and their derivatives are bounded on $z\in(0,y+A)$.  
Using this notation the result after doing the $K$ integral can be expressed in compact form, where the integrands can be expressed using factors $l_1(z), l_1(\alpha), l_2(\rho)\cdots$ etc.
Further use of results from appendix \ref{Bsmooth}, l'H\^opital's rule and Lemma~\ref{lemma-a} renders
\bea
\lim_{y+A\to 1_-}\mathcal{G}^-_{\rm out} = \lim_{y+A\to 1_-}\left[2\ln \left(\frac{w-y-2A}{w-y}\right)-2 \text{Ci}(w-y-2A)+2 \text{Ci}(w-y)\right]\,.\nonumber\\ \label{Gsmallres}
\eea
Next we consider $\mathcal{G}^+_{\rm out}$. 
Interchanging the order of integration and performing the $K$ integrals we obtain
\bea
\mathcal{G}^+_{\rm out} &=& 
\frac{2}{B}\int^{y+A}_0 dz\, B'(z) \text{Ci}(\alpha )  +
\frac{2}{B}\int^{y+A}_0 dz\, B'(z) (\text{Ci}(\chi )-\text{Ci}(\beta )-\text{Ci}(\xi ))\nonumber \\[4mm]
&-&\frac{4}{B^2} 
\int^{y+A}_0 dx \,B'(x)\int^{x}_0 dz\, B'(z) 
\text{Ci}(\tau)
\nonumber \\[4mm]
&+&
\frac{2}{B^2} 
\int^{y+A}_0 dx \,B'(x)\int^{x}_0 dz\, B'(z) 
(\text{Ci}(\rho )+\text{Ci}(\sigma ))\,.\nonumber\\
\label{kdoneGlarge}
\eea
We pass the lengthy procedures of carefully manipulating the double integrals using 
Lemma~\ref{lemma-a} and l'H\^opital and point out that the nonvanishing terms in $\mathcal{G}^+_{\rm out}$ cancel out the cosine integrals in (\ref{Gsmallres}). Final result is
\bea
\label{Gout-morefinal}
\lim_{y+A\to 1_-}{\cal G}_{\rm out} &= & \lim_{y+A\to 1_-}\left[2\ln\left(\frac{w-y-2A}{w-y}\right)\right]\,
\eea
which is in agreement with (\ref{after-T00}).


\section{Inside region}
\label{inside-region}
In this section we look at the energy density in the region $r<a$. 
The idea is to see if there is a symmetry between the inside and outside regions that would allow us to extract the final result for the inside region from the results we have already calculated which are valid outside the shell.  
We consider each of the four pieces: $F_-$, $F_+$, ${\cal G}^-$ and ${\cal G}^+$.


We have already outlined how to calculate $F_-^{out}$ in the outside region, doing the same calculation in the inside region we find out
%
\bea
\label{in-out-1}
F^{\rm in}_-(y,w)\big|_{A\to-A} - F^{\rm out}_-(y,w)=0\,.
\eea
Here, $F^{\rm in}_-(y,w)\big|_{A\to-A}$  is the inside expression transformed by making sign change on $A\to -A$ and switches $(y,w)\to (w,y)$. A similar expression holds for ${\cal G}^-_{\rm in}(w,y)$ through transformation $(y,w,A)\to (w,y,-A)$
\bea
\label{in-out-3}
{\cal G}_{\rm in}^-(w,y)|_{A\to-A} - {\cal G}_{\rm out}^-(y,w) =0 \,.
\eea
Defining positive definite variables throughout the inside region
\bea
\hat\alpha && = -A+w-z\,,
\hat\gamma = A+w-z\,,
\hat\beta = 3 A-w+z\,,\nonumber\\
\hat\sigma &&= 2 A+x-z\,,
\hat\rho = 2 A-x+z\,,\nonumber
\label{exps-in-1}
\eea   
we transform the inside result again  and find 
\bea\label{in-out-2}
&& F^{\rm in}_+(w,y)\big|_{A\to-A} - F^{\rm out}_+(y,w)=\text{imaginary}\,.
\eea
With additional definitions $\hat\chi = A+y-z$, $\hat\eta = A-y+z$, we obtain
\bea\label{in-out-4}
 && {\cal G}_{\rm in}^+(w,y)\big|_{A\to-A} - {\cal G}_{\rm out}^+(y,w) =\text{imaginary}\,.
\eea
From equations (\ref{in-out-1}, \ref{in-out-3}, \ref{in-out-2},  \ref{in-out-4}) we see that the energy density inside the shell can be obtained from the outside results by performing the transformation $(y,w,A)\to (w,y,-A)$ and dropping any imaginary parts that are produced.

\end{document}